\documentclass[11pt,aps,floats,amssymb,tightenlines,nofootinbib]{revtex4}
\usepackage{color}
\usepackage{graphicx}
\usepackage{amsmath}
\usepackage{amsfonts}
\usepackage{amscd}
\usepackage{epsfig}
\usepackage{amssymb}
\usepackage{tabularx}
\usepackage{soul}

\begin{document}

\title{Junction conditions for higher order gravity theories from a Gibbons-Hawking-York boundary term}
\author{Marcos A. Ramirez${}^1$, Cristi\'an Mart\'{i}nez${}^{2,3}$}
\affiliation{${}^1$ Instituto de Astronom\'ia Te\'orica y Experimental, CONICET - Universidad Nacional de C\'ordoba, (5000) C\'ordoba, Argentina \\ ${}^2$Centro de Estudios Cient\'ificos (CECs), Av. Arturo Prat 514, Valdivia, Chile\\
${}^3$Facultad de Ingenier\'{\i}a, Universidad San Sebasti\'an, General Lagos 1163, Valdivia, Chile
}

\begin{abstract}
    In this work we study the problem of generalizing the Gibbons-Hawking-York boundary terms for general quadratic theories of gravity and propose a simple condition to obtain them. 
    From these terms we derive the junction conditions for a subset of this family of theories that includes Gauss-Bonnet (GB) gravity. We re-obtain the well-known results for GB theory, generalize them to other quadratic theories and compare the resulting junction conditions with the ones already derived in the literature using other methods.    
\end{abstract}

\maketitle

\section{Introduction}

Higher order in curvature gravity theories have always been
a natural extension to General Relativity (GR). They are interesting from a classical perspective since the study of their properties can reveal the uniqueness or naturalness of GR among the possible metric theories of gravity and provide a natural comparison framework when assessing experimental tests of GR. They also appear naturally as classical limits in theoretical frameworks that seek to unify all fundamental interactions. For these theories the field equations are typically of fourth order in the metric, which make the search for solutions with minimally coupled matter-energy fields a highly non-trivial task \cite{Belenchia}.

Thin shells are a physically reasonable simplification for describing matter fields. It may allow to find analytical solutions that incorporate matter into theories that are very complex in full generality. There are several examples of theories of gravity to which matter solutions are not currently known and thin shells can be a way to get an intuition about the role that matter plays in these theories. 

Thin shell solutions are typically found by solving junction conditions. Roughly speaking, there are in the literature two types of derivations for the junction conditions of a given theory that correspond to different traditions in theoretical and mathematical physics. These two approaches would typically give different answers but they do coincide in GR. One approach, related to methods in mathematical physics, is based on a careful analysis of tensor distributions as weak solutions of the field equations. In this approach the field equations are the main objects, it is about making sense out of these equations when curvature is represented by a tensor distribution. One of the most comprehensive papers for $f(R)$ theories is the one by Senovilla \cite{Senovilla13} and for quadratic theories the one by Reina et. al. \cite{RSV}. We can also mention \cite{Rosa}, where this approach is extended to gravity theories with extra scalar degrees of freedom (which includes Lagrangians made of arbitrary functions of the three quadratic invariants constructed out of the Riemann tensor), and \cite{ChuTan}, where they propose a generalization of the allowed distributional objects but it is also in tension with previous works.

The other approach has more to do with a field-theoretical tradition. It consists on adding an appropriate boundary term to the action whose variation, when coupled with matter fields on the boundary, would naturally give junction conditions: the so called Gibbons-Hawking-York (GHY) boundary term. In this approach the field equations are not the starting point, instead, the junction conditions are derived together with the field equations by extremizing an action, namely, by requiring that its variation vanishes. Since boundary terms naturally appear in this variation, in order to attain an extreme for the action that do not require extra boundary conditions (independent from the field equations), it is necessary to add a suitable boundary action whose variation exactly cancels out them.  
Junction conditions are in this approach on an equal footing with field equations. This was the method adopted by Myers \cite{Myers} and Davis \cite{Davis} in order to derive the most frequently used Einstein-Gauss-Bonnet (EGB) junction conditions. More recently, it was also used to derive these conditions for general Lovelock theories \cite{Chakraborty}. 

There are a number of issues for both methods when trying to apply them to general quadratic theories. In order to make sense out of distributional solutions of the field equations one is forced to impose conditions on the extrinsic curvature of the shell originating from restrictions on the operations between distributions. This restrictions make the construction fundamentally different from what can be done in GR. On the other hand, there are also conceptual issues when one tries to find the GHY terms for quadratic theories other than GB \cite{DyerHinterbichler}. The natural boundary terms of the variation of the action are not always integrable, and one should impose extra assumptions to make them so. 

In this article a method to obtain the Gibbons-Hawking-York boundary terms for general quadratic theories of gravity is presented. For the cases in which we can find this term we also derive the junction conditions and compare them with previous results. In section \eqref{nbtQ} we derive the natural boundary terms that arise from the variations of the Lagrangians. Then in section \eqref{GHYbt} we derive the GHY boundary terms by imposing a simple prescription on the variation. In section \eqref{secJC}, for the theories whose GHY terms can be found by this method, we derive the corresponding junction conditions. Finally, in \eqref{comparison} we compare with previous results and make final comments. Our conventions for the Riemann tensor and extrinsic curvature can be summarized as follows
\begin{eqnarray}
R^\alpha_{\beta \mu \nu}v^{\beta}&=&(\nabla_{\mu}\nabla_{\nu}-\nabla_{\nu}\nabla_{\mu})v^{\alpha} \nonumber \\
K_{ij}&=&(e_i)^{\mu}(e_j)^{\nu}\nabla_{\mu}n_{\nu} \nonumber,
\end{eqnarray}
where greek indexes denote bulk coordinates, latin indexes surface coordinates, $v^{\alpha}$ is an arbitrary vector field, $(e_i)^{\mu}$ are surface coordinate basis vectors and $n^{\alpha}$ is the normal vector field of the surface\footnote{With the notation $K_{\alpha \beta}$ we mean $\nabla_{\alpha}n_{\beta}$, where the normal field $n^{\alpha}$ is geodesically extended out of the surface. All normal derivatives of the extrinsic curvature are being applied to $\nabla_{\alpha}n_{\beta}$.}.

\section{Natural boundary terms for quadratic gravity} \label{nbtQ}

In this paper we will consider the three possible independent theories constructed out of second order products of the Riemann tensor. The independent Lagrangians are:

\begin{eqnarray}
    {\cal L}_1 &=& R^2 \\ {\cal L}_2&=&R^{\alpha\beta}R_{\alpha\beta}  \\  {\cal L}_3&=&R^{\alpha\beta\mu\nu}R_{\alpha\beta\mu\nu}. 
\end{eqnarray}
The variation of these Lagrangians can be calculated using the following relations
\begin{eqnarray}
\label{deltaR}
\delta R^{\alpha}_{\beta\mu\nu} &=& \nabla_\mu (\delta \Gamma^{\alpha}_{\beta\nu}) - \nabla_\nu (\delta \Gamma^{\alpha}_{\beta\mu}) \\
\label{deltaGamma}
\delta\Gamma^{\alpha}_{\mu\nu}&=& \frac{1}{2} g^{\alpha\beta}(\nabla_\mu \delta g_{\beta\nu} + \nabla_\nu \delta g_{\beta\mu}-\nabla_\beta \delta g_{\mu\nu}) 
\end{eqnarray}

Taking into account these identities, and using both integration by parts and the Bianchi identities, one can obtain the following relations

\begin{eqnarray}
\frac{1}{\sqrt{-g}}\delta ({\sqrt{-g}} {\cal L}_1) &=& 2\left( \nabla^2 R g_{\alpha\beta}-\nabla_\alpha \nabla_\beta R + R R_{\alpha \beta}-\frac{1}{4}R^2g_{\alpha\beta} \right) \delta g^{\alpha \beta} + \nonumber \\ 
&&  + 2 \nabla_{\mu} \left( R g^{\alpha\beta} \delta\Gamma^{\mu}_{\alpha\beta} - R g^{\alpha\mu} \delta\Gamma^{\beta}_{\alpha\beta}  + (\nabla^{\mu}R g^{\alpha\beta} - \nabla^{\alpha}R g^{\mu\beta}) \delta g_{\alpha\beta} \right) \label{v1} \\ 
\frac{1}{\sqrt{-g}}\delta ({\sqrt{-g}} {\cal L}_2) &=& \left(\frac{1}{2} g_{\alpha\beta}(\nabla_{\mu}\nabla^{\mu} R-R_{\mu\nu}R^{\mu\nu})-2R_{\beta\mu\nu\alpha}R^{\mu\nu}-\nabla_{\alpha}\nabla_{\beta} R + \nabla_{\mu}\nabla^{\mu}R_{\alpha\beta}\right) \delta g^{\alpha\beta} + \nonumber \\ 
&& + \nabla_{\mu}\left(2R^{\alpha\beta}\delta\Gamma^{\mu}_{\alpha\beta} -2R^{\alpha\mu}\delta\Gamma^{\nu}_{\alpha\nu} +\left(\frac{1}{2}g^{\alpha\beta}\nabla^{\mu}R -2\nabla^{\alpha}R^{\mu\beta} +\nabla^{\mu}R^{\alpha\beta}\right)\delta g_{\alpha\beta}\right) \label{v2}\\
\frac{1}{\sqrt{-g}}\delta ({\sqrt{-g}} {\cal L}_3) &=&  \left(2R_{\gamma\mu\nu\alpha}{R^{\gamma\mu\nu}}_{\beta}-\frac{1}{2}g_{\alpha\beta}R_{\gamma\eta\mu\nu}R^{\gamma\eta\mu\nu}+4\nabla_{\mu}\nabla_{\nu}{{{R^{\mu}}_{\alpha}}^{\nu}}_{\beta}\right) \delta g^{\alpha\beta} + \nonumber \\
&& +4\nabla_{\mu}\left({R_{\alpha}}^{\beta\mu\nu}\delta\Gamma^{\alpha}_{\beta\nu}-\nabla_{\nu}R^{\alpha\mu\nu\beta}\delta g_{\alpha\beta}\right). \label{v3}
\end{eqnarray}

In these expressions one can recognize both the tensors that must be proportional to the stress-energy tensor in the field equations and the natural boundary terms for each theory. From the boundary terms one might be able to find the respective Gibbons-Hawking-York boundary actions under the Dirichlet condition (fixed induced metric on the boundary). We will prove in this Section, as it happens in GR, that the variation of any of the quadratic actions can be written as follows

\begin{equation}
{\delta \cal I}={\delta \cal F}+{\delta \cal B} =  {\delta \cal F} +\int_{\partial M} \sqrt{|h|}{\cal B}_h^{ij} \delta h_{ij} + \int_{\partial M}\sqrt{|h|} {\cal B}_K^{ij} \delta K_{ij}, \label{variation} 
\end{equation}
where ${\delta \cal F}$ stands for the bulk component of the variation while ${\delta \cal B}$ is the emergent boundary component, the latter being the integral on $\partial M$ of an expression that contains a term that involves a contraction with $\delta h_{ij}$ and another term with $\delta K_{ij}$. 
This last term is the one that one seeks to cancel out with an additional Gibbons-Hawking-York term, which can be defined from its variation as

\begin{equation}
\label{GHYdef}
{\delta \cal I}_{GHY}\equiv- \int_{\partial M} \sqrt{|h|}{\cal B}_K^{ij} \delta K_{ij} .
\end{equation}

In order to identify the coefficients ${\cal B}_h^{ij}$ and ${\cal B}_K^{ij}$ we set a normal coordinate system centered on the boundary. We assume that the boundary is composed by different sections that are either timelike or spacelike (see Fig.\ref{fig1}) connected by zero-measure sections where the normal vector is ill-defined. For each of these (non zero-measure) sections we can write the metric ansatz in a normal neighbourhood     
\begin{equation}
ds^2 = \epsilon d\eta^2 + h_{ij} (\eta,x) dx^i dx^j,     
\end{equation}
where $\epsilon$ is $1$ ($-1$) for the timelike (spacelike) sections. With this ansatz one can write the different components of $\delta \Gamma^{\mu}_{\alpha\beta}$ as follows.
\begin{eqnarray}
\delta \Gamma^{\eta}_{i j} &=&-\epsilon \delta K_{ij} \\
\delta \Gamma^{i}_{\eta j} &=& \delta h^{ik} K_{kj} + h^{ik}\delta K_{kj}\\
\label{deltagamma}
\delta \Gamma^{i}_{jk}&=&\frac{1}{2} h^{im}(D_j\delta h_{km} + D_k\delta h_{jm}-D_m\delta h_{jk})
\end{eqnarray}
where $K_{ij}$ is the extrinsic curvature (in our convention equivalent to $\frac{1}{2} \partial_{\eta} h_{ij}$) and $D$ denotes the covariant derivative compatible with $h_{ij}$.

Using an extended version of the Gauss-Codazzi identities for writing the different components of the full Riemann tensor in terms of the intrinsic metric, its associated (intrinsic) Riemann tensor, the extrinsic curvature and their first derivatives in $\eta$, and assuming the Dirichlet condition ($\delta h_{ij}=0$), we can express the variations of the emergent boundary terms by giving the coefficients

\begin{eqnarray}
{\cal B}^{ij}_{K1}& = &-4 R h^{ij} = -4\left(\hat{R}-\epsilon(2\partial_{\eta}K+K_{lm}K^{lm}+K^2) \right) h^{ij}\label{b1} \\
{\cal B}^{ij}_{K2}& = &  -2 \left(\hat{R}^{ij}-\epsilon((\partial_{\eta}K+K^{lm}K_{lm})h^{ij}+2K^{ik}K_k^j+KK^{ij}+\partial_{\eta}K^{ij}) \right)   \label{b2} \\
{\cal B}^{ij}_{K3} &= & 8 \epsilon\left(\partial_{\eta}K^{ij}+3K^i_m K^{mj} \right) , \label{b3}
\end{eqnarray}
where the hats indicate intrinsic (boundary) curvature tensors. From these expressions we can eventually find suitable GHY terms, and after we find these boundary terms we can derive the junctions conditions by considering the full variations. The variations of the boundary terms under Neumann boundary conditions ($\delta K_{ij}=0$) are complementary to the ones under Dirichlet condition above and can be expressed with the coefficients   
\begin{eqnarray}
\label{h1}
{\cal B}_{h1}^{ij}&=& 2(
RK^{ij}+\partial_{\eta}Rh^{ij}) \\ 
\label{h2}
{\cal B}_{h2}^{ij}&=& \frac{1}{2} \partial_{\eta}R h^{ij} +\partial_{\eta}R^{ij} + 4K^{(i}_kR^{j)k} +  \epsilon\left(2R^{\eta\eta}K^{ij}+D_k(R^{\eta k})h^{ij} - 2D^i (R^{\eta j})\right) \\
\label{h3}
{\cal B}_{h3}^{ij} &=& 4\left(\epsilon\left(\partial_{\eta}R^{\eta i \eta j}+
2K^{(i}_kR^{j)\eta k \eta} + K R^{\eta i \eta j} + 2D_k(R^{\eta i kj})\right)+ K_{kl}R^{ilkj}\right), 
\end{eqnarray}
where $D()$ denotes that the object between parenthesis is to be understood as an intrinsic tensor on the boundary. Both ${\cal B}_{h2}^{ij}$ and ${\cal B}_{h3}^{ij}$ are obtained taking into account a divergence that can be understood as a ``boundary of boundary" term and that is to be discarded, in both cases related to the $\delta \Gamma^i_{jk}$ coefficients. The complete expressions for these boundary variations are 
\begin{eqnarray}
{\delta \cal B}_2&=& \int_{\partial M} \sqrt{-\epsilon h} \left( {\cal B}_{h2}^{ij}\delta h_{ij} +{\cal B}_{K2}^{ij}\delta K_{ij} - \epsilon D_k\left(R^{\eta k}h^{ij}\delta h_{ij}\right) \right) \\
{\delta \cal B}_3&=& \int_{\partial M} \sqrt{-\epsilon h} \left( {\cal B}_{h3}^{ij}\delta h_{ij} +{\cal B}_{K3}^{ij}\delta K_{ij} + 4\epsilon D_k\left(R^{\eta ijk}\delta h_{ij}\right)\right),
\end{eqnarray}
from which one can recognize the contribution of the divergences in both (\ref{h2}) and (\ref{h3}) by taking into account their construction by means of integration by parts of the derivatives in (\ref{deltagamma}).
It is noteworthy that this construction is not necessary for ${\delta \cal B}_1$ because of the absence of $\delta \Gamma^i_{jk}$ in (\ref{v1}). In this way, the coefficients ${\cal B}_{h}^{ij}$ and ${\cal B}_{K}^{ij}$ fully characterize the boundary variations for the three Lagrangians. 

In the same way as in (\ref{K1}, \ref{K2}, \ref{K3}), the ${\cal B}_{h}^{ij}$ coefficients can also be written in terms of $K_{ij}$, $\partial_\eta K_{ij}$, $D_kK_{ij}$, $h_ij$ and the intrinsic Riemann tensor on the boundary by using the following generalized Gauss-Codazzi relations
\begin{eqnarray}
\label{dic1}
R^{\eta i \eta j} &=& -\partial_{\eta} K^{ij} - 3K^i_mK^{mj} \\
\label{dic2}
R^{ijk \eta} &=&\epsilon (D^iK^{jk}-D^jK^{ik}) \\
\label{dic3}
R^{ijkl} &=& \hat{R}^{ijkl}+2\epsilon K^{l[i}K^{j]k} \\
\label{dic4}
R^{ij} &=& \hat{R}^{ij} -\epsilon (\partial_\eta K^{ij}+2K^{ik}K^j_k + KK^{ij}) \\
\label{dic5}
R^{i\eta} &=&\epsilon(D_jK^{ij}-D^iK) \\
\label{dic6}
R^{\eta\eta}&=&-\partial_{\eta}K - K_{ij}K^{ij} \\
\label{dic7}
R&=&\hat{R}-\epsilon(2\partial_{\eta}K+K_{ij}K^{ij}+K^2).
\end{eqnarray}  
The final expressions after making all the replacements are quite lengthy and not very illuminating, and will be used only when needed hereafter.

\section{Gibbons-Hawking-York boundary terms} \label{GHYbt}

The main difficulty associated with this method arises when one tries to integrate (\ref{GHYdef}). One should find an expression defined on the boundary that is a function of $h_{ij}$, $K_{ij}$ and $\partial_{\eta}K_{ij}$ such that its variation exactly cancels out the $\delta K_{ij}$ part of the natural boundary term. However, as we will see in this Section, in general one can find the suitable expressions only if one is allowed to control the variation of the normal derivatives of the extrinsic curvature. The necessity of restricting the variational problem in order to find the GHY terms has already been acknowledged in the literature, but with different proposed resolutions. For example, in \cite{DyerHinterbichler} for $f(R)$ theories (which includes $R^2$ theory) it is proposed to fix the bulk scalar curvature in the boundary. A similar restriction for more general non-linear in $R$ theories was proposed in \cite{BalcerzakDbrowski}. Likewise, in \cite{Berezin} an arbitrary (but fixed) linear relation between $\delta h_{ij}$ and $\delta K_{ij}$ is imposed. In this work we will adopt a different approach and propose what we think is the simplest, yet geometrically meaningful, prescription: we fix the covariant normal derivative of the extrinsic curvature. Although we believe that any of these procedures should be better justified, it is likely possible to impose additional boundary data beyond just the intrinsic metric without exhausting the solution space of the field equations, as these equations are generally of fourth order\footnote{The initial plus boundary value problem for quadratic theories is, to our knowledge, still an open problem.}.

More precisely, every term in the coefficients ${\cal B}_K^{ij}$ of the natural boundary expressions (\ref{b1}, \ref{b2}, \ref{b3}) is either of second order in $K_{ij}$, first order in $\partial_{\eta}K^{ij}$ or first order in the intrinsic Riemann tensor. The GHY Lagrangian densities must then be composed of terms either of third order in $K^{ij}$ or proportional to (some contraction of) $K^{ij}\partial_{\eta}K^{kl}$ or $\hat{R}^{ijkl}K^{mn}$, but there are two different problems with this procedure. On the one hand, not every possible variation of the form (with a given contraction) $K^{ij}K^k_l\delta K_{mn}$ can be obtained with the variation of a third order expression in $K_{ij}$. On the other hand, the variation of any expression containing $K^{ij}\partial_{\eta}K^{kl}$ would necessarily have a term involving $\delta(\partial_{\eta} K_{ij})$ that is not present in any of the natural boundary terms.

To understand the first of these problems, consider the four independent $(2,0)$ tensors constructed out of a second order expression in $K_{ij}$ and the intrinsic metric: 
\begin{equation}
\label{KK}
    K^2h^{ij} \;,\; K^{lm}K_{lm}h^{ij} \;,\;  KK^{ij} \;,\; K^{ik}K^j_k.
\end{equation} 
These four tensors appear in the different versions of ${\cal B}_K^{ij}$ (\ref{b1}, \ref{b2}, \ref{b3}), but there are only three possible independent scalars of third order in $K_{ij}$ whose variations are
\begin{eqnarray}
 \delta(K^3) &=& 3K^2\delta K = 3K^2 (-K^{ij}\delta h_{ij} + h^{ij}\delta K_{ij})  \label{K1} \\
 \delta(K^{ij}K_{ij} K) &=& -(K_{lm}K^{lm}K^{ij}+2KK^{ik}K_k^j) \delta h_{ij} +(K_{lm}K^{lm}h^{ij}+2KK^{ij})\delta K_{ij}  \label{K2} \\
\delta (K^{ij}K_{jk}K^k_i) &=& - 3K^{ik}K^{jl}K_{kl}\delta h_{ij}+3K^i_kK^{jk}\delta K_{ij} \label{K3}. 
\end{eqnarray}
Under the Dirichlet condition, two of the four tensors in (\ref{KK}), $K^{lm}K_{lm}h^{ij}$ and $KK^{ij}$, appear in the variation of only one scalar: $K_{ij}K^{ij}K$, so if Eq. (\ref{GHYdef}) is to be integrable then the relative coefficient for these two tensors must be the same as in (\ref{K2}). That is, the variation should contain a term proportional to $(K_{lm}K^{lm}h^{ij}+2KK^{ij})\delta K_{ij}$ as an integrability condition. There are two different Lagrangians whose ${\cal B}_K^{ij}$ coefficients contain at least one of these second order in $K$ tensors: (\ref{b1}) contains $4K_{lm}K^{lm}h^{ij}$ and (\ref{b2}) contains $2K_{lm}K^{lm}h^{ij}+2KK^{ij}$. That means that among the possible Lagrangians
\begin{equation}
{\cal L}= \alpha {\cal L}_1 + \beta {\cal L}_2 + \gamma {\cal L}_3,
\end{equation}
{\it the ones whose boundary variations are integrable in this sense are those that satisfy} $\beta=-4\alpha$. Thus, in this article we will derive the GHY boundary terms and the corresponding junction conditions for theories that satisfy this property, which can be naturally classified into theories where $\alpha=\beta=0$ and $\gamma\neq0$ (which we will call Kretschmann theory), $-4\alpha=\beta\neq 0$ and $\gamma=0$, and $-4\alpha=\beta\neq 0$ and $\gamma\neq 0$. Note that Gauss-Bonnet theory is included in this last class: $\alpha=\gamma=-\frac{1}{4}\beta$.

The second of these problems is more subtle, as it implies that one should reduce the variational problem in order to get well-defined GHY terms. One might even simply appeal to cancel out all the $K^{ij}\partial_{\eta}K^{kl}$ terms \cite{Bunch}, which is a very restrictive hypothesis. If one imposes the cancellation of these terms the only quadratic theory left turns out to be Gauss-Bonnet, which can be checked from the expressions (\ref{b1}), (\ref{b2}) and (\ref{b3}). As mentioned, the alternative that we are going to consider in this work is to control the variation of higher-order derivatives of $K_{ij}$ in the normal direction. 
This is an ad-hoc restriction on the space of boundary data for the theory which fundamentally change the variation problem, allowed by the four-order nature of the resulting field equations. There are several possible and nonequivalent ways to do this as one does not have the same variational problem whether one imposes $\delta(\partial_{\eta}K_{ij})=0$ or $\delta(\partial_{\eta}K^{ij})=0$ or if one fixes any other of the higher-order ordinary partial derivatives in the normal direction. One of the simplest and geometrically meaningful conditions that one might propose is $\delta(n^{\mu}\nabla_{\mu}K_{\alpha\beta})=0$ \footnote{It can be shown that all normal components ($\alpha=\eta$ or $\beta=\eta$) vanish for this quantity, so there is no ambiguity in writing $n^{\mu}\nabla_{\mu}K_{ij}$ when referring to the same object.}, which is the one we will adopt here. It has the advantage of being a covariant prescription, so  
one can get rid of the arbitrariness in the index structure of the condition that is present when imposing a fixed value of an ordinary derivative. 

After writing this covariant derivative in Gaussian coordinates and the $\eta ij$ components of the Christoffel symbols in terms of the extrinsic curvature, the variation of the normal derivative of $K_{ij}$ can be expressed as follows
\begin{equation}
\label{partialetavar}
  \delta(\partial_{\eta}K_{ij})= \delta(n^{\mu}\nabla_{\mu}K_{ij}) + 4K^k_{(i}\delta K_{j)k}-2K^k_{(i}K^l_{j)}\delta h_{kl},
\end{equation}
for which our prescription simply means the vanishing of the first term of the RHS. Using this relation and the expressions (\ref{K1}, \ref{K2}, \ref{K3}) we can find the GHY terms for the families of theories described above.

We point out two important properties that make this possible. First, the $\partial_{\eta}K$ in the ${\cal B}_K^{ij}$ coefficients cancel out exactly, so there is no need to calculate $\delta(\partial_{\eta} K)$.
To illustrate this point consider the natural boundary variation under Dirichlet condition for the $(\alpha=-\frac{1}{4}\beta=1 , \gamma=0)$ theory
\begin{equation}
\label{1-4variation}
{\cal B}_{K(1,-4,0)}^{ij}=-4\left(\epsilon(2\partial_{\eta}K^{ij}+4K^{ik}K^j_k+2KK^{ij}+(K^{lm}K_{lm}-K^2)h^{ij})-2\hat{G}^{ij}\right),
\end{equation}
where $\hat{G}^{ij}$ represents the intrinsic Einstein tensor. It is explicit that both the $\partial_{\eta} K$ term is canceled out and the aforementioned integrability condition gets satisfied.

Second, when trying to find an expression whose variation cancels out terms proportional to $\partial_{\eta}K^{ij}\delta K_{ij}$ it is necessary to compute $K^{ij}\delta(\partial_{\eta}K_{ij})$. Under our prescription (and Dirichlet condition), $K^{ij}\delta(\partial_{\eta}K_{ij})$ will only contribute through a $K^i_kK^{jk}\delta K_{ij}$ term (via the second term of the RHS of (\ref{partialetavar})), which is easily integrable by (\ref{K3}). This second property is particularly important when comparing to other possible reductions of the variational problem: (\ref{partialetavar}) do not interfere with the integrability condition that stems from the second order in $K^{ij}$ terms of ${\cal B}_K^{ij}$. If $K^{lm}K_{lm}h^{ij}\delta K_{ij}$ or $KK^{ij}\delta K_{ij}$ appeared in (\ref{partialetavar}) then this integrability condition would be affected when considering the full variation and it would change the family of theories for which the GHY term can be derived through this approach. Our choice has then the non-trivial feature that the integrability condition that arises from the $K^{ij}K^{kl}\delta K_{mn}$ terms in the natural boundary variations turns out to be independent from the prescription on $\delta(n^{\mu}\nabla_{\mu}K_{\alpha \beta})$. 

In this way, the expression that we are going to use to find the GHY terms under our hypothesis is the following
\begin{equation}
\label{deltaKdK}
\delta (K^{ij}\partial_{\eta}K_{ij}) = K^{ij}\delta(n^{\mu}\nabla_{\mu}K_{ij}) + (\partial_{\eta}K^{ij}+8K^{ik}K^j_k)\delta K_{ij} - 2(K^{(i}_k\partial_{\eta}K^{j)k}+5K^{ik}K^{jl}K_{kl}) \delta h_{ij},
\end{equation}
for which our prescription, as in (\ref{partialetavar}), simply implies the vanishing of the first term of the RHS.

\subsection{Kretschmann theory}

In order to find the GHY action for this theory we use the following ansatz 
\begin{equation}
\label{GHansatz}
    {\cal I}_{GH3}= -8\epsilon\int_{\partial M} \sqrt{-\epsilon h} \left(K^{ij}\partial_{\eta}K_{ij} -C K^{ij}K_{jk}{K^k}_i\right),
\end{equation}
and then vary this expression in order to find the constant $C$. Taking into account (\ref{deltaKdK})
and imposing $\delta(n^{\mu}\nabla_{\mu}K_{\alpha\beta})=0$, one can write the variation of (\ref{GHansatz}) under the Dirichlet condition as
\begin{equation}
\label{GHansatzvar}
\delta{\cal I}_{GHY3} = - 8\epsilon \int_{\partial M} \sqrt{-\epsilon h} \left(\partial_{\eta}K^{ij}+(8-3C)K^{ik}K^j_k\right)\delta K_{ij}. 
\end{equation}
Since ${\cal B}_{K3}^{ij}=8\epsilon(\partial_{\eta}K^{ij}+3K^{ik}K^j_k)$, in order to satisfy (\ref{GHYdef}) we should set $C=5/3$. In this way, for the ${\cal L}_3$ theory the Gibbons-Hawking-York boundary term is
\begin{equation}
\label{GH3}
    {\cal I}_{GHY3}=8\epsilon\int_{\partial M} \sqrt{-\epsilon h}  \left(\frac{5}{3}K^{ij}K_{jk}{K^k}_i-K^{ij}\partial_{\eta}K_{ij}\right).
\end{equation}

\subsection{$R^2-4R_{\alpha\beta}R^{\alpha\beta}$ theory}
In this case the variation that one seeks to cancel with the GHY term is the one defined by (\ref{1-4variation}). The GHY action can be calculated by taking into account (\ref{K1}, \ref{K2}) for the $K^2h^{ij}\delta K_{ij}$ and $(2KK^{ij}+K^{lm}K_{lm})h^{ij}\delta K_{ij}$ terms respectively, and the variation of (\ref{GHansatz}), given by (\ref{GHansatzvar}), 
for the $\partial_{\eta}K^{ij}\delta K_{ij}$ and $K^{ik}K^j_k \delta K_{ij}$ terms. Since these last two terms appear in ${\cal B}_K^{ij}$ as $\partial_{\eta}K^{ij}+2K^{ik}K^j_k$, we should set $C=2$ in (\ref{GHansatz}) (such that $8-3C=2$), so the GHY action for this theory finally reads     
\begin{equation}
\label{GH1-4}
    {\cal I}_{GHY}=4\int_{\partial M} \sqrt{-\epsilon h} \left(\epsilon\left(2(K^{ij}\partial_{\eta}K_{ij}-2K^{ij}K_{jk}{K^k}_i)+K_{ij}K^{ij} K - \frac{1}{3}K^3\right) - 2\hat{G}^{ij}K_{ij} \right).
\end{equation}

\subsection{Gauss-Bonnet theory}

For this particular theory the resulting GHY boundary action depends only on $K_{ij}$ and $h_{ij}$ as all terms containing $\partial_{\eta} K_{ij}$ cancel out in the combination 
\begin{equation}
    {\cal B}^{ij}_{K-GB} =  {\cal B}^{ij}_{K1}-4{\cal B}^{ij}_{K2}+{\cal B}^{ij}_{K3}.
\end{equation}
As mentioned, neither of the two problems that emerge when trying to find the GHY action appear in this case, so it can be found just by applying Eqs. (\ref{K1}, \ref{K2}, \ref{K3}), or by adding the RHSs of (\ref{GH3}) and (\ref{GH1-4}), namely
\begin{equation}
\label{GHGB}
{\cal I}_{GHY-GB}= 4 \int_{\partial M} \sqrt{-\epsilon h} \left( \epsilon\left(K^{ij}K_{ij}K-\frac{1}{3}K^3-\frac{2}{3}K^{ij}K_{jk}K^k_i\right)-2 \hat{G}^{ij}K_{ij} \right).
\end{equation}
This expression coincides with the one found in \cite{Davis} for $\epsilon=1$.

\subsection{General $-4\alpha=\beta$ theories}

Gauss-Bonnet theory is just a particular case of this family. Being all variations linear operators, for arbitrary $\alpha$ and $\gamma$ the general GHY term would be
\begin{eqnarray}
\label{GHY}
{\cal I}_{GHY}&=& 4 \int_{\partial M} \sqrt{-\epsilon h} \left(\epsilon\left(2(\alpha-\gamma)K_{ij}\partial_{\eta}K^{ij} + \frac{2}{3}\left(5\gamma-6\alpha\right) K^{ij}K_{jk}K^k_i\right) \right. \nonumber + \\
&& \left.+\alpha\left(\epsilon\left((K^{ij}K_{ij})K-\frac{1}{3}K^3\right)-2 \hat{G}^{ij}K_{ij}\right)\right).
\end{eqnarray}
This expression includes all the particular cases (\ref{GH3}), (\ref{GH1-4}) and (\ref{GHGB}).

\section{Junction conditions}  \label{secJC}

Having derived the GHY term for any quadratic theory satisfying condition $-4\alpha=\beta$ (\ref{GHY}), under the prescription of fixing the normal covariant derivative of  $K_{ij}$, we can now proceed to compute the junction conditions as follows.

\begin{figure}
    \centering
    \includegraphics[width=0.3\linewidth]{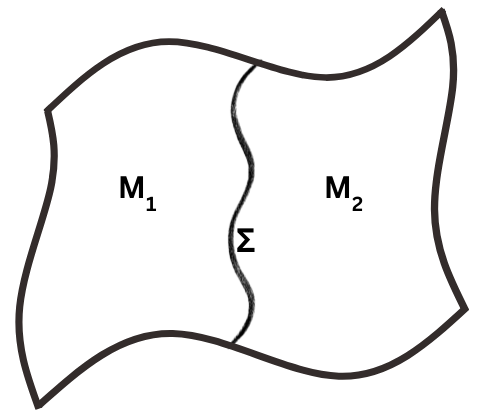}
    \caption{Schematic representation of the spacetime in which the variational problem is defined. The boundaries of each region have four sections: two spacelike and two timelike. The identification is made in one of the timelike sections, all other three sections of each individual boundary constitute the boundary of the composite manifold $\partial(M_1\cup M_2)$.}
    \label{fig1}
\end{figure}

As explained in the Introduction, this derivation consists of writing a suitable action principle for the problem at hand. A thin shell here is characterized as the intersection of the boundaries of two different manifolds, that together represent two non-intersecting regions of the bulk spacetime, as illustrated in Fig. \ref{fig1}. In general, this intersection is only a part of the boundaries and can be thought of as a pair of subsets of each boundary that are mutually identified (here denoted $\Sigma_i$ when treated as part of the boundary of any of the bulk regions or $\Sigma$ when treated as a manifold on its own). The parts of the boundaries not being identified constitute the boundary of the composite manifold $M_1\cup M_2$, and as such would typically have both timelike and spacelike sections. On the other hand, $\Sigma$ will be assumed to be timelike in order to represent the world tube of eventual matter-energy fields. The identification is made such that the first fundamental form is unambiguously defined, i.e. continuity of the metric at this common boundary. We will also assume sufficient regularity in the manifold structures such that a Gaussian coordinate system (extending into both manifolds) can be defined in a neighborhood of this shared boundary. 

It would be convenient to write the GHY action in terms of a Lagrangian density as
\begin{equation}
{\cal I}_{GHY}=\int_{\partial M} \sqrt{|h|} {\cal L}_{GHY}.
\end{equation}
The total action, including possible matter-energy fields described by ${\cal L}_{m}$, defined within the boundary, can be then written as follows
\begin{equation}
\label{action}
{\cal I} = \int_{M_1 \cup M_2} \sqrt{-g} {\cal L} + \int_{\partial M_1\cup\partial M_2}\sqrt{|h|} {\cal L}_{GHY}+ \int_{\Sigma}\sqrt{-h} {\cal L}_{m},
\end{equation}
where $\partial M_1 \cup \partial M_2$ includes $\Sigma$ ``twice'', in the sense that involves the evaluation of the integrals at both sides of the orientable surface, and the boundary of the composed manifold $\partial(M_1 \cup M_2)$. ${\cal L}_{GHY}$ should have the same functional form at both sides with respect to the metric and their derivatives, but these integrals do not necessarily coincide since the normal derivatives (of any order) of $h_{ij}$ can be different at each side. Thus, the variation of this action for the theories considered in this paper is
\begin{eqnarray}   
 \delta{\cal I} &=& \int_{M_1 \cup M_2} \frac{\delta(\sqrt{-g}{\cal L})}{\delta g_{\mu\nu}} \delta g_{\mu\nu} - \frac{1}{2}\int_{\Sigma} \sqrt{-h} S^{ij} \delta h_{ij} + \nonumber \\ && \int_{\Sigma_1\cup\Sigma_2} \left( \left( \sqrt{-h}{\cal B}_h^{ij} + \frac{\delta(\sqrt{-h}{\cal L}_{GHY})}{\delta h_{ij}}  \right) \delta h_{ij} + \frac{\delta(\sqrt{-h}{\cal L}_{GHY})}{\delta (n^{\mu}\nabla_{\mu}K_{ij})} \delta (n^{\mu}\nabla_{\mu}K_{ij})\right) + \nonumber \\ &&   \int_{\partial(M_1\cup M_2)} \left( \left( \sqrt{|h|}{\cal B}_h^{ij} + \frac{\delta(\sqrt{|h|}{\cal L}_{GHY})}{\delta h_{ij}}  \right) \delta h_{ij} + \frac{\delta(\sqrt{|h|}{\cal L}_{GHY})}{\delta (n^{\mu}\nabla_{\mu}K_{ij})}  \delta (n^{\mu}\nabla_{\mu}K_{ij})\right), 
\end{eqnarray}
where $\Sigma_1$ and $\Sigma_2$ are the parts of $\partial M_1$ and $\partial M_2$, respectively, that are being identified, and $S^{ij}\equiv-2\delta {\cal L}_m/\delta h_{ij}-{\cal L}_m h^{ij}$. By construction, the natural boundary $\delta K_{ij}$ terms have been canceled by the GHY actions. The terms that are evaluated at the boundary of $M_1\cup M_2$ must be addressed in exactly the same way as in a single manifold with boundary setting, and hence they are unrelated to the thin shell construction so their contribution is not going to be considered in this work. They can be dismissed by some form of asymptotic prescription, or by simply fixing all relevant degrees of freedom at this outer boundary. Since, in our approach, $\delta(n^{\mu}\nabla_{\mu}K_{ij})$ is to be suppressed, all variations in $\Sigma$ can be finally written in terms of $\delta h_{ij}$. In this way, a stationary action gives junction conditions on the boundary of the form
\begin{equation}
\label{junction}
\left[{\cal B}_h^{ij} + \frac{\delta{\cal L}_{GHY}}{\delta h_{ij}}+\frac{1}{2}{\cal L}_{GHY}h^{ij}\right]=\frac{1}{2}S^{ij},
\end{equation}
where the brackets denote the difference between the values that these quantities acquire when evaluated at each of the sides. The subtraction is due to the fact that the definition of all quantities inside the bracket depend on the orientation of the normal vector when applying Stoke's theorem to the divergences in Eqs. (\ref{v1}), (\ref{v2}) and (\ref{v3}). The integrands of the natural boundary terms come from contractions of some (variational) vector quantity with the normal vector, and when applying the identification between $\Sigma_1$ and $\Sigma_2$ the associated normal vectors must be opposite to each other ($n^{\mu}_1=-n^{\mu}_2$). Then the integrands for the natural boundary terms must have opposite signs when computing the integral over $\Sigma$, and the same should happen with the GHY variations since they are defined in terms of the natural boundary terms (\ref{GHYdef}). The positive sign in the bracket is then given by the (arbitrary) orientation of the Gaussian normal coordinate.

It is important to note that the variations of normal derivatives of $K_{ij}$ are solely contained in the variation of GHY action, they are not present in the natural boundary terms. This means that they are not a fundamental property of the theory but rather a byproduct of the method, hence the plurality of non-equivalent ways of treating them is not surprising.

\subsection{Kretschmann theory}

We are now able to compute Eq. (\ref{junction}) for this theory. The natural boundary coefficients for $\epsilon=1$  (\ref{h3}) can be written in terms of the two fundamental forms and derivatives of the extrinsic curvature by applying Eqs. (\ref{dic1}, \ref{dic2}, \ref{dic3}) and (\ref{K3}) as follows
\begin{eqnarray}
\label{h3f}
  {\cal B}_{h3}^{ij} &=& 4\left(-\partial^2_{\eta}K^{ij}-8K^{(i}_m \partial_{\eta}K^{j)m}-K\partial_{\eta}K^{ij}-K^{ik}K^{jl}K_{kl}+K^{ij}K^{lm}K_{lm}-3KK^{im}K_m^j + \right. \nonumber \\ && \left. +K_{kl}\hat{R}^{iklj}+ 2(D_kD^{(i}K^{j)k}-D_kD^kK^{ij}) \right),    
\end{eqnarray}
where the symmetrization come from reorganizing ${\cal B}_{h3}^{ij}\delta h_{ij}$ taking into account the symmetry of the intrinsic metric.    

We then compute the variation of the GHY term (\ref{GH3}) from (\ref{K3}) and (\ref{deltaKdK}) and obtain its contribution to the junction conditions as follows
\begin{equation}
\label{GH3var}
\frac{\delta {\cal L}_{GHY3}}{\delta h_{ij}}+\frac{1}{2}{\cal L}_{GHY3}h^{ij}=4\left(4K^{(i}_m \partial_{\eta}K^{j)m}-14K^{ik}K^{jl}K_{kl} +\left(-K_{lm}\partial_{\eta}K^{lm}+\frac{5}{3}K^{lm}K_{lk}K^k_m\right)h^{ij}\right).
\end{equation}
In this way, according to the proposed method, the junction conditions  for the Kretschmann theory are given by
\begin{eqnarray}
\label{JC3}
\frac{1}{8\gamma} S^{ij} &=& -[\partial^2_{\eta}K^{ij}] - 4[K_m^{(i}\partial_{\eta} K^{j)m}]-[K\partial_{\eta} K^{ij}]-[K_{lm}\partial_{\eta} K^{lm}]h^{ij}+[-15K^{im}K^{jl}K_{lm}+K^{lm}K_{lm} K^{ij} - \nonumber \\ &&-3KK_m^iK^{jm}]+\frac{5}{3}[K^{lm}K_l^kK_{mk}]h^{ij} + \hat{R}^{iklj}[K_{kl}]+2D_kD^{(i}[K^{j)k}]-2D_kD^k[K^{ij}].
\end{eqnarray}
These conditions are very different in nature to the Israel-Darmois junction conditions or to the EGB junction conditions obtained in \cite{Davis}, mainly because of the presence of both normal and intrinsic derivatives of the extrinsic curvature. In particular, as we will discuss in the last section, they can support concentrated matter-energy fields even if $K_{ij}$ is continuous across $\Sigma$.   

\subsection{$R^2-4R_{\alpha\beta}R^{\alpha\beta}$ theory}

For this theory, the natural boundary coefficients that are part of the junction conditions are a combination of (\ref{h1}) and (\ref{h2}) as follows.
\begin{eqnarray}
\label{h1-4}
{\cal B}_{h(1,-4,0)}^{ij}&=&2(R-4R^{\eta\eta})K^{ij}-4\left(-2D^i R^{\eta j}+\partial_{\eta}R^{ij} + 4K^{(i}_kR^{j)k} + (D_kD_lK^{kl}-D_kD^kK) h^{ij}\right) \nonumber \\ &=&4\left(\partial^2_{\eta}K^{ij}+8K^{(i}_m \partial_{\eta}K^{j)m}+K\partial_{\eta}K^{ij}-\frac{1}{2}(K_{lm}K^{lm}+K^2)K^{ij}+4KK^i_mK^{mj}- \right. \nonumber \\ && \left. -2K^{(i}_k\hat{R}^{j)k}-2K_{kl}\hat{R}^{iklj}+\frac{1}{2}\hat{R}K^{ij}+D_kD^kK^{ij}-D^iD^jK+(D_kD^kK-D_kD_lK^{kl})h^{ij}\right) \nonumber\\&&
\end{eqnarray}
To get this expression we used the Gauss-Codazzi relations with $\epsilon=1$ (\ref{dic4}, \ref{dic5}, \ref{dic6} and \ref{dic7}) in order to replace the different Ricci bulk components and bulk scalar curvature, with the exception of $\partial_{\eta}R^{ij}$ \footnote{We avoided an assessment of this term through (\ref{dic4}) since finding an expression of $\partial_{\eta}\hat{R}^{ij}$ in terms of $K^{ij}$ and its derivatives is a more involved calculation.}, that was obtained indirectly through a suitable contraction of the differential Bianchi identity as follows. The normal derivative of the Ricci tensor comes from the term $n_{\mu}\nabla^{\mu}R^{ij}\delta h_{ij}$, that is present in the natural boundary term of the $R_{\mu\nu}R^{\mu\nu}$ theory (\ref{b2}), and the following expression that come from explicit evaluation in Gaussian coordinates 
\begin{equation}
\label{etaricci}
 n_{\mu}\nabla^{\mu}R^{ij}=\partial_{\eta}R^{ij}+2K^{(i}_kR^{j)k} .
\end{equation}
In turn, (\ref{etaricci}) can be written in terms of other derivatives (and suitable contractions) of the Riemann tensor by means of the differential Bianchi identity  
\begin{equation}
\label{bianchi}
  n_{\mu}\nabla^{\mu}R^{ij}\delta h_{ij}=(n_{\mu}(\nabla^{i} R^{\mu j}-\nabla_{\nu}R^{i\mu\nu j})) \delta h_{ij}.
\end{equation}
The right hand side of this expression, which consists of terms already present in (\ref{h2}) and (\ref{h3}), does not contain $\partial_{\eta}R^{ij}$ and was applied to derive (\ref{h1-4}).

There is another subtlety regarding the expression (\ref{h1-4}). The RHS of (\ref{bianchi}), after applying (\ref{dic2}) and (\ref{dic5}), involves different contractions of the object $D_iD_jK^{kl}$. We applied the intrinsic Riemann tensor on the second order tensor $K^{ij}$ to reverse the order of the derivatives in one of the resulting terms. In addition, we discarded an explicitly antisymmetric expression involving second intrinsic derivatives of the extrinsic curvature, since, when contracted with $\delta h_{ij}$, a symmetric object, it makes no contribution to the natural boundary variation.     

On the other hand, the variation of the GHY Lagrangian (\ref{GH1-4}) with $\epsilon=1$ can be calculated from (\ref{K1}), (\ref{K2}) and (\ref{deltaKdK}). In order to compute $K^{ij}\delta\hat{G}_{ij}$ one should resort to intrinsic versions of (\ref{deltaR}, \ref{deltaGamma}), and, after two integrations by parts, it can be shown that it can be expressed as combinations of second intrinsic covariant derivatives of $K^{ij}$ contracted by $\delta h_{ij}$. The variation of ${\cal L}_{GHY}$ turn out to be          
\begin{eqnarray}
\label{GH1-4var}
\frac{\delta{\cal L}_{GHY(1,-4,0)}}{\delta h_{ij}}&=&4\left(-4K^{(i}_m \partial_{\eta}K^{j)m}+16K^{im}K^{jl}K_{lm}-K^{lm}K_{lm} K^{ij}-2KK_m^iK^{jm}+K^2K^{ij}-\right. \nonumber\\
&&-\hat{R}K^{ij}-K\hat{R}^{ij}+4\hat{R}^{k(i}K^{j)}_k-2D_kD^{(i}K^{j)}_k+D_kD^kK^{ij}+D^iD^jK+\nonumber\\ &&\left.+(D_kD_lK^{kl}-D_kD^kK)h^{ij}\right).
\end{eqnarray}

We can now compute (\ref{junction}) for this theory from (\ref{h1-4}) and (\ref{GH1-4var}) as follows
\begin{eqnarray}
\label{JC1-4}
\frac{1}{8\alpha}S^{ij} &=& [\partial^2_{\eta}K^{ij}] + 4[K_m^{(i}\partial_{\eta} K^{j)m}]+[K\partial_{\eta} K^{ij}]+[K_{lm}\partial_{\eta} K^{lm}]h^{ij} +\left[16K^{ik}K^{jl}K_{kl}-\frac{3}{2}K_{lm}K^{lm}K^{ij}+\right. \nonumber \\ && \left. +\frac{1}{2}K^2K^{ij}+2KK^i_mK^{jm}\right]+\left[\frac{1}{2}K_{lm}K^{lm}K-\frac{1}{6} K^3-2K^{lm}K_{mk}K^k_l\right]h^{ij}- \nonumber \\&& -[K_{kl}]\left(\hat{R}^{iklj}+\hat{P}^{iklj}\right)-2D_kD^{(i}[K^{j)k}]+2D_kD^k[K^{ij}],
\end{eqnarray}
where $\hat{P}^{ijkl}$ is the divergence-free part of the intrinsic Riemann tensor, which can be defined as
\begin{equation}
\hat{P}^{ijkl}=\hat{R}^{ijkl}+2\hat{R}^{j[k}h^{l]i}-2\hat{R}^{i[k}h^{l]j}+\hat{R}h^{i[k}h^{l]j}.
\end{equation}
Like (\ref{JC3}), this junction conditions are not merely algebraic in $K_{ij}$ at the surface but also involve the jump of both normal and intrinsic derivatives of the extrinsic curvature.

\subsection{Gauss-Bonnet theory}
Every term in (\ref{junction}) is linear in the Lagrangian of theory. The GHY term, as defined by (\ref{GHYdef}), also depends linearly on the Lagrangian since variations are linear operators. We can then derive the junction conditions for the $\alpha=\gamma=-\frac{1}{4}\beta$ case by simply adding (\ref{JC3}) and (\ref{JC1-4}) as follows
\begin{eqnarray}
\label{JCGB}
   \frac{1}{8\alpha}S^{ij} &=& \left[K^{ik}K^{jl}K_{kl}-\frac{1}{2}K_{lm}K^{lm}K^{ij}+\frac{1}{2}K^2K^{ij}-KK^i_mK^{jm}\right]+\left[\frac{1}{2}K_{lm}K^{lm}K-\frac{1}{6} K^3- \right. \nonumber \\
   && \left. \frac{1}{3}K^{lm}K_{mk}K^k_l\right]h^{ij}-\hat{P}^{iklj}[K_{kl}], 
\end{eqnarray}
which, after taking into account the different definitions of the coupling constants, is the expression obtained in \cite{Davis}. It is noteworthy that all terms involving derivatives of $K^{ij}$ from (\ref{JC3}) are canceled out by (\ref{JC1-4}).

\subsection{General $-4\alpha=\beta$ theories}
By combining (\ref{JC3}) and (\ref{JC1-4}) we can now express the most general junction conditions of this work 
\begin{eqnarray}
\label{JCmain}
   \frac{1}{8}S^{ij} &=& (\alpha-\gamma)\left([\partial^2_{\eta}K^{ij}] + 4[K_m^{(i}\partial_{\eta} K^{j)m}]+[K\partial_{\eta} K^{ij}]+[K_{lm}\partial_{\eta} K^{lm}]h^{ij} +2D_kD^k[K^{ij}] -\right. \nonumber \\ && \left. -2D_kD^{(i}[K^{j)k}]-\hat{R}^{iklj}[K_{kl}] \right)+\alpha \left(\frac{1}{2}[K^2K^{ij}] +\left(\frac{1}{2}[K_{lm}K^{lm}K]-\frac{1}{6} [K^3]\right)h^{ij}-\right. \nonumber \\ && \left. -\hat{P}^{iklj}[K_{kl}] \right) +(16\alpha-15\gamma)[K^{ik}K^{jl}K_{kl}]+\frac{1}{2}(2\gamma-3\alpha)[K_{lm}K^{lm}K^{ij}]+ \nonumber \\ && + (2\alpha-3\gamma)[KK^i_mK^{jm}]+\frac{1}{3}(5\gamma-6\alpha)[K^{lm}K_{mk}K^k_l]h^{ij},
\end{eqnarray}
where (\ref{JC3}), (\ref{JC1-4}) and (\ref{JCGB}) are all particular cases.

\section{Comparison with previous results and final comments} \label{comparison}

One of the main differences between this result and \cite{RSV} or \cite{ChuTan}, and potentially any other work in the distributional approach, is that the junction conditions here derived do not involve normal components of the distributional stress-energy tensor. In our work this is a consequence of varying quantities defined on the shell (first and second fundamental forms) and not the surface itself or the normal vector in particular. This is related to a more profound difference between the junction conditions in \cite{RSV} and the ones of this work and of \cite{ChuTan}: the appearance of a double layer term, which enforces the existence of a gravitational ``dipole moment density'' on the thin shell. Since the structure of the junction conditions turns out to be very different, in order to make a comparison between these three manuscripts we should restrict ourselves to tangential components and to the $\delta_{\Sigma}$ (Dirac´s delta over $\Sigma$) terms of \cite{RSV}.

Another important difference between this work and both \cite{RSV} and \cite{ChuTan} is the necessity of the Lichnerowicz condition, i.e. the continuity of the extrinsic curvature at the shell. In \cite{RSV} it is argued that in order to interpret the thin shell construction as a well-defined distributional solution of quadratic gravity one should avoid $\delta_{\Sigma}^2$-terms. It is then shown that the only way to avoid them in general quadratic theories is to impose the Lichnerowicz condition ($R^2$ theory being the only exception). In the case of \cite{ChuTan} the construction is different, they resort to a regularization scheme in order to make sense out of distributional products in the field equations beyond the classical theory of distributions. But for the theories we are considering in this paper, the only way to obtain definite (i.e. not dependent on an arbitrary smearing function) junction conditions is either by imposing the Lichnerowicz condition or by imposing $[K^{ij}][K_{ij}]=0$.     

In this way, in order to make a comparison between a part of the junction conditions of these references, i.e. the tangential part within the $\delta_{\Sigma}$ term, and our junction conditions (\ref{JCmain}) we will first apply the Lichnerowicz condition. The resulting conditions are then
\begin{equation}
\label{JCLich}
 \frac{1}{8}S^{ij} = (\alpha-\gamma)\left([\partial^2_{\eta}K^{ij}] + 4K_m^{(i}[\partial_{\eta} K^{j)m}]+K[\partial_{\eta} K^{ij}]+K_{lm}[\partial_{\eta} K^{lm}]h^{ij}\right).
\end{equation}

On the other hand, in \cite{RSV} the coupling constants are grouped into two factors: $\kappa_1=2\alpha+\beta/2$ and $\kappa_2=2\gamma+\beta/2$. In our theories we should set $\kappa_1=0$ and $\kappa_2=2(\gamma-\alpha)$, so Eq. 51 of \cite{RSV} in our notation can be written as
\begin{equation}
\label{JCRSV}
\frac{1}{4}\kappa S^{ij}=(\alpha-\gamma)\left([\partial^2_{\eta}K^{ij}] + 4K_m^{(i}[\partial_{\eta} K^{j)m}]+K[\partial_{\eta} K^{ij}]\right),
\end{equation}
where $\kappa$ is present because of a different definition of the coupling constants related to the matter-energy Lagrangian. In our notation the gravitational coupling constant is incorporated into $\alpha$ and $\gamma$, which is equivalent to set $2\kappa=1$ (there is no $1/2\kappa$ factor in front of the gravitational terms of (\ref{action})). It is then remarkable that the resulting conditions, being obtained by means of a completely different method, are very similar to ours. The difference being the term $K_{lm}[\partial_{\eta} K^{lm}]h^{ij}$, which comes from the variation of the volume element within the GHY action. It might not be surprising that such a term does not appear in approaches that do not apply the variation of a supplemented action.   

If we assume the lack of double layers, Eq. 113 of \cite{RSV}, another restriction is enforced. In our notation it can be written as $[\partial_{\eta}K]=0$. Notably, this assumption also implies that all non-tangential terms within $\delta_{\Sigma}$ vanish. In this way, if we further assume the continuity of $\partial_{\eta}K_{ij}$, our junction conditions fully coincide with the ones of \cite{RSV}: it would be no double layer, no normal components in the $\delta_{\Sigma}$ term, and the first terms of the RHSs of both (\ref{JCLich}) and (\ref{JCRSV}) match perfectly.      

In the same way, Eq. 80 of \cite{ChuTan}, after adapting the expression to our notation, reads
\begin{equation}
\frac{1}{4}\kappa S^{ij}= (\gamma-\alpha)\left(K_m^{(i}[\partial_{\eta} K^{j)m}]-K[\partial_{\eta} K^{ij}]+K^{ij}[\partial_{\eta}K]\right),
\end{equation}
where $\kappa$ stands for $8\pi$ in their convention (again, absorbed into $\alpha$ and $\gamma$ in our notation, and it should be set to $1/2$ for a comparison). These conditions are very different from ours and from the ones derived in \cite{RSV}. It is noteworthy that, although their method is more similar to the one used in \cite{RSV} than to our method (based on distributional solutions of the established field equations rather than supplementing the action of the theory itself), their results for the theories considered in this paper is much more divergent.

Summarizing, in this work we derived a method to obtain the Gibbons-Hawking-York term for a family of quadratic theories under the prescription $\delta(n^{\mu}\nabla_{\mu}K_{ij})=0$. This prescription has the property that, while being among the simplest that one might write, it does not alter the integrability condition for the $K^{ij}K^{kl}\delta K_{mn}$ terms within the natural boundary terms. We were able to find the GHY term (\ref{GHY}) for theories satisfying $\beta=-4\alpha$ and recovered, for Gauss-Bonnet gravity, the GHY term obtained in \cite{Davis}. We then applied this term to derive the junction conditions for this family of theories (\ref{JCmain}) and also recovered the Gauss-Bonnet results from \cite{Davis}. Finally, after imposing continuity of the extrinsic curvature at the shell, we compared the resulting junction conditions to other ones obtained through other methods. Our results fully coincide with \cite{RSV} under the special case $[\partial_{\eta} K_{ij}]=0$.    

\vspace{2mm}
\noindent{\bf{Acknowledgments}}\\[1mm]
We thank Ernesto Eiroa for insightful comments, questions and suggestions. We also thank Osvaldo Moreschi and Felipe Mena for useful discussions. This research has been partially funded by ANID FONDECYT grants 1220862 and 1241835. MAR is funded by CONICET.

\bibliography{main.bib}

\end{document}